\documentclass{aastex63}

\usepackage{bm}

\received{-----}
\revised{March 13, 2021}
\accepted{April 6, 2021}
\submitjournal{ApJ}

\shorttitle{Sample article}
\shortauthors{Yoshida and Kokubo}

\graphicspath{{./}{figures/}}

\begin{document}

\title{ELEMENTARY PROCESS OF GALACTIC SPIRAL ARM FORMATION: PHASE SYNCHRONIZATION OF EPICYCLIC MOTION BY GRAVITATIONAL SCATTERING}

\correspondingauthor{Yoshida Yuki}
\email{yoshida-yuki520@g.ecc.u-tokyo.ac.jp}

\author[0000-0002-2631-7095]{Yuki Yoshida}
\affiliation{Department of Astronomy, The University of Tokyo, 7-3-1 Hongo, Bunkyo-ku, Tokyo 113-0033, Japan}

\author[0000-0002-5486-7828]{Eiichiro Kokubo}
\affiliation{Center for Computational Astrophysics, National Astronomical Observatory of Japan, 2-21-1 Osawa, Mitaka, Tokyo 181-8588, Japan}

\begin{abstract}

Swing amplification is a model of spiral arm formation in disk galaxies. 
Previous $N$-body simulations show that the epicycle phases of stars in spiral arms are synchronized. 
However, the elementary process of the phase synchronization is not well understood. 
In order to investigate phase synchronization,  we investigate the orbital evolution of stars due to gravitational scattering by a perturber under the epicycle approximation and its dependence on orbital elements and a disk parameter.
We find that gravitational scattering by the perturber can cause phase synchronization of stellar orbits.
The epicycle phases are better synchronized for smaller initial epicycle amplitudes of stars and larger shear rates of galactic disks.
The vertical motion of stars does not affect the phase synchronization.
The phase synchronization forms trailing dense regions, which may correspond to spiral arms.

\end{abstract}
\keywords{galaxies: kinematics, galaxies: spiral, method: numerical simulation}

\section{Introduction} \label{sec:intro}

Spiral structures are a common pattern for astrophysical disks. 
Disk galaxies with spiral arms are called spiral galaxies.
Grand-design spiral galaxies have a few global, continuous, symmetric spiral arms; multi-arm spiral galaxies have an intermediate-scale spiral structure; and flocculent spiral galaxies have many intangible irregular local spiral arms.
Some protoplanetary gas disks show spiral arms that may be produced by gravitational instability \citep{2016Sci...353.1519P}. 
It is also theoretically suggested that Saturn's rings have small spiral arms or wakes,
considered to be formed due to the gravitational instability of the ring.

The formation of galactic spiral arms is an important astrophysical problem.
The formation models can be divided into two classes: those with and without perturbers.
Gravitational interaction of the disk with a perturber leads to spiral arm formation.
Perturbers can be internal, such as bars\citep{2015MNRAS.454.2954B}, or external, such as satellite galaxies\citep{2016MNRAS.458.3990P}.
Spiral arms develop spontaneously in disks without perturbers.
There are two formation models for spontaneous spirals: density waves and material arms.
In the density wave model, spiral arms are considered to be quasi-stationary standing patterns \citep{1964ApJ...140..646L}. 
The spiral arms are periodic coarse waves of the surface density that propagate through the disk.
However, no simulations show stationary spiral arms.
In the material arm model, stars gather due to their self-gravity to form spiral arms.
The spiral arms are transient and recurrent \citep{1989MNRAS.240..991S}, as observed in $N$-body simulations \citep[e.g.,][]{1984ApJ...282...61S, 2000Ap&SS.272...31S, 2009ApJ...706..471B, 2011ApJ...730..109F}. 
When spiral arms recur, swing amplification is considered to be the formation mechanism.
In swing amplification, the surface number density of stars is enhanced when Toomre's $Q \simeq 1$--$2$, and a leading mode of the stellar distribution is amplified and begins to trail during rotation\citep{1965MNRAS.130..125G, 1966ApJ...146..810J, 1981seng.proc..111T}.

Some studies have confirmed that swing amplification is appropriate for spiral formation.
\cite{2014ApJ...787..174M} performed local $N$-body simulations to investigate the relation between the pitch angle of spiral arms and the disk parameter, and showed that swing amplification can explain this relation.
\cite{2016ApJ...821...35M} calculated the pitch angle,  wavelength, and amplification factor as a function of the disk parameters based on swing amplification, and performed local $N$-body simulations to compare with their formula.
They found that the results of the simulation are consistent with their formula.
They also estimated the number of spiral arms based on swing amplification.
\cite{2018MNRAS.481..185M} performed global $N$-body simulations of spiral galaxies and found that the pitch angle and number of spiral arms agree with the formula of \cite{2014ApJ...787..174M} and \cite{2016ApJ...821...35M}.  
Furthermore, the observed pitch angle agrees with those in the $N$-body simulations and the swing amplification model\citep[e.g.,][]{2006ApJ...645.1012S, 2019ApJ...871..194Y}.

Studies have been undertaken to investigate the mechanism of swing amplification itself.
\cite{1966ApJ...146..810J} showed that a perturber in a disk forms a spiral structure by gravitational interaction. 
The gravity of the perturber induces an in-phase epicyclic motion of stars that leads to the formation of a spiral-like dense pattern. 
\cite{2016ApJ...823..121M} derived the perturbation equation of the epicyclic motion of stars and investigated the process of swing amplification in detail.
They found that the epicycle phases are synchronized during density amplification.
However, the mechanism of  the phase synchronization and its dependence on orbital elements and disk parameters have not yet been clarified.

In this paper we study the orbital evolution of stars due to scattering by a perturber under the epicycle approximation and clarify the synchronization process of epicyclic motion. 
In Section \ref{sec:sys-sim}, the model and simulation method are described. 
Section \ref{sec:result} presents the results of the simulation and Section \ref{sec:discussion} is devoted to a discussion.
Finally, Section \ref{sec:summary} summarizes the paper.

\section{Model and Method} \label{sec:sys-sim}

\subsection{Epicycle Approximation} \label{sub:epicycle-approximation}

We use the epicycle approximation for stellar motion in a galactic disk \citep[e.g.,][]{1987gady.book.....B, 1992PASJ...44..601K}. 
We adopt a local Cartesian coordinate system $(x, y, z)$ whose origin rotates around the galactic center with the circular frequency $\Omega$,  where the $x$-axis is directed radially outward, the $y$-axis is parallel to the direction of rotation, and the $z$-axis is normal to the $x$--$y$ plane. 
We introduce a perturber at the coordinate origin.
The equation of motion of a star is given by
\begin{equation} \label{eq:motion}
\left\{
\begin{array}{l}
\ddot{x} = 2\Omega\dot{y} + (4\Omega^2-\kappa^2)x -\displaystyle\frac{GM_{\rm{p}}}{(r^2+c^2)^{3/2}}x,\\
\ddot{y} = -2\Omega\dot{x} - \displaystyle\frac{GM_{\rm{p}}}{(r^2+c^2)^{3/2}}y,\\
\ddot{z} = -\nu^2 z - \displaystyle\frac{GM_{\rm{p}}}{(r^2+c^2)^{3/2}}z,
\end{array}
\right.
\end{equation}
where $M_{\rm{p}}$ is the pertureber mass, $G$ is the gravitational constant, $c$ is the softening parameter, and $\kappa$ and $\nu$ are the epicycle and vertical frequencies given by 
\begin{eqnarray} \label{eq:frequency}
\left\{
\begin{array}{l}
\kappa^2 \equiv \left(R\displaystyle\frac{{\rm d}\Omega^2}{{\rm d}R} + 4\Omega^2\right)_{(a,0,0)}, \\
\nu^2 \equiv \left(\displaystyle\frac{\partial^2 \Phi}{\partial z^2}\right)_{(a,0,0)}.
\end{array}
\right.
\end{eqnarray}
In Equation (\ref{eq:motion}), $2\Omega\dot{y}$ and $-2\Omega\dot{x}$ are the Coriolis force, $4\Omega^2 x$ is the centrifugal force, and $-\kappa^2 x$ and $-\nu^2 z$ represent the galactic gravitational force. 
The last terms in Equation (\ref{eq:motion}) represent the gravity of  the perturber. 

We can normalize time by $\Omega^{-1}$ and length by the tidal radius of the perturber given by
\begin{equation} \label{eq:Hill}
r_{\rm{t}} \equiv \left[\frac{M_{\rm{p}}}{(4-\kappa^2/\Omega^2) M_{\rm{g}}} \right]^{1/3} a,
\end{equation}
where $a$ is the galactocentric distance and $M_{\rm{g}}$ is the effective mass of the galaxy that satisfies $GM_{\rm{g}}/a^2=\Omega^2a$.
Then the normalized form of Equation (\ref{eq:motion}) is 
\begin{equation} \label{eq:normalized-motion}
\left\{
\begin{array}{l}
\ddot{\tilde{x}} = 2\dot{\tilde{y}} + \alpha\tilde{x} -\displaystyle\frac{\alpha}{(\tilde{r}^2+\tilde{c}^2)^{3/2}}\tilde{x},\\
\ddot{\tilde{y}} = -2\dot{\tilde{x}} - \displaystyle\frac{\alpha}{(\tilde{r}^2+\tilde{c}^2)^{3/2}}\tilde{y},\\
\ddot{\tilde{z}} = -\beta^2 \tilde{z} - \displaystyle\frac{\alpha}{(\tilde{r}^2+\tilde{c}^2)^{3/2}}\tilde{z},
\end{array}
\right.
\end{equation}
where $\alpha\equiv4-\kappa^2/\Omega^2$ and $\beta\equiv\nu/\Omega$. 
Hereafter, variables with tildes on top are normalized. 

Neglecting the gravity of the perturber we can derive the analytical solution to Equation (\ref{eq:normalized-motion}) as
\begin{eqnarray} \label{eq:solution}
\left\{
\begin{array}{l}
\tilde{x} = \tilde{b} - \tilde{a}_{\rm e} \cos\left(\gamma \tilde{t}-\phi_{\rm e} \right),\\
\displaystyle\tilde{y} = \tilde{y}_{\rm c} - \frac{\alpha}{2}\tilde{b}\tilde{t} + 2\frac{\tilde{a}_{\rm e}}{\gamma} \sin\left(\gamma \tilde{t}-\phi_{\rm e} \right),\\
\tilde{z} = \tilde{a}_{\rm v} \sin\left(\beta \tilde{t}-\phi_{\rm v} \right),
\end{array}
\right.
\end{eqnarray}
and the velocity components as
\begin{eqnarray} \label{eq:solution-velocity}
\left\{
\begin{array}{l}
    \dot{\tilde{x}} = \gamma \tilde{a}_{\rm e} \sin\left(\gamma \tilde{t}-\phi_{\rm e} \right),\\
    \displaystyle\dot{\tilde{y}} = -\frac{\alpha}{2}\tilde{b} + 2\tilde{a}_{\rm e} \cos\left(\gamma \tilde{t}-\phi_{\rm e} \right),\\
    \dot{\tilde{z}} = \beta \tilde{a}_{\rm v} \cos\left(\beta \tilde{t}-\phi_{\rm v} \right),
\end{array}
\right.
\end{eqnarray}
where $\gamma = \kappa/\Omega$ and $\tilde{a}_{\rm e}, \tilde{a}_{\rm v}, \tilde{y}_{\rm c}, \tilde{b}, \phi_{\rm e}, \phi_{\rm v}$ are integration constants: $\tilde{a}_{\rm e}$ and $\phi_{\rm e}$ are the amplitude and phase of epicyclic motion, $\tilde{a}_{\rm v}$ and $\phi_{\rm v}$ are those of vertical oscillation, and $b$ and $\tilde{y}_{\rm c}$ are the impact parameter and $\tilde{y}$ of the guiding center at $\tilde{t}=0$.
The period of the epicyclic motion is $2\pi/\gamma$ in the normalized units. 

Equation (\ref{eq:normalized-motion}) includes the Jacobi integral:
\begin{equation} \label{eq:jacobi}
\tilde{E}_{\rm{J}} = \frac{\gamma^2 \tilde{a}_{\rm e}^2}{2} + \frac{\beta^2 \tilde{a}_{\rm v}^2}{2} - \frac{\tilde{b}^2}{8}\alpha(4-\alpha) - \frac{\alpha}{(\tilde{r}^2+\tilde{c}^2)^{1/2}}.
\end{equation}
We numerically integrate Equation (\ref{eq:normalized-motion}) using the Runge--Kutta--Fehlberg method.
Here, we take $\tilde{c}=1.0$ in the simulation.

\subsection{Simulation Models} \label{sub:models}

\subsubsection{Disk Parameters} \label{subsub:disk-parameters}

The galactic disk rotates differentially with a shear rate $\Gamma$ given by
\begin{equation} \label{eq:shear}
\Gamma \equiv -\frac{{\rm d} \log \Omega}{{\rm d} \log R} = 2 - \frac{\gamma^2}{2} = \frac{\alpha}{2}.
\end{equation}
The shear rate ranges from 0 for a rigid rotation to 3/2 for Kepler rotation.
The observed frequencies in the solar neighborhood are $\gamma=1.39$ and $\beta=3.81$\citep{1987gady.book.....B}. 
For $\gamma=1.39$, the shear rate is $\Gamma\simeq1.034$.
We take $\Gamma=1.034$ mainly, and use $\Gamma=0.5, 1.0, 1.5$ to compare each case.
To investigate the dependence on the shear rate, we vary $\Gamma$ from 0 to 1.5.
The vertical frequency is fixed to $\beta=3.81$.

\subsubsection{Initial Conditions}
\label{subsub:initial-condition}

The initial position and velocity of a star  $(\tilde{x},\tilde{y},\tilde{z},\dot{\tilde{x}},\dot{\tilde{y}},\dot{\tilde{z}})$ are given from the initial orbital elements $(\tilde{a}_{\rm e, ini}, \tilde{a}_{\rm v, ini}, \tilde{y}_{\rm c}, \tilde{b}, \phi_{\rm e, ini}, \phi_{\rm v,ini})$.
We introduce the subscript ``0" to represent the orbital elements of a star at $\tilde{y}=\infty$, and the subscript ``ini" to represent the orbital elements when we start the numerical integration.
To investigate the dependence on the epicyclic and vertical oscillations, the initial epicycle amplitude $\tilde{a}_{\rm e,0}$ is varied between 0 and 2.0, and the vertical amplitude $\tilde{a}_{\rm v,0}$ is varied as 0.125 and 0.500.
The epicycle phase $\phi_{\rm e,0}$ is basically prepared as $\{0,\pi/32,\ \cdots,63\pi/32\}$, and the initial vertical phase $\phi_{\rm v,0}$ is prepared as $\{0,\pi/8,\ \cdots,15\pi/8\}$.
We take $\tilde{y}_{\rm c}=0$, which means that $\tilde{y}$ of the guiding center is located on $\tilde{x}$-axis at $\tilde{t}=0$.
We also take the impact parameter $\tilde{b}_0$ to be 2.5 to 6.0 to investigate the dependence on the impact parameter.
Next, we set the initial orbital element based on the above parameters.
We assume that the epicyclic and vertical oscillations do not change before the initial condition: $\tilde{a}_{\rm e, ini}=\tilde{a}_{\rm e,0}$, $\tilde{a}_{\rm v, ini}=\tilde{a}_{\rm v,0}$, $\phi_{\rm e,ini}=\phi_{\rm e,0}$, and $\phi_{\rm v,ini}=\phi_{\rm v,0}$.
The initial impact parameter $\tilde{b}_{\rm{ini}}$ can be calculated based on the conservation of the Jacobi integral of Equation (\ref{eq:jacobi}).
We also set the initial $\tilde{y}$ of the guiding center to be $-\alpha \tilde{b}_{\rm ini}\tilde{t}_{\rm ini}/2=32.0$.
The integrating time of the simulation is $\tilde{t}=32.0$.

Next, we simulate a group of stars to investigate a distribution.
We take $\tilde{y}_{\rm c}=0$ for all the stars.
We prepare two cases for the initial epicycle phases: for one case $\phi_{\rm e,0}$ is determined randomly, while for the other $\phi_{\rm e,0}=0, \pi/2, \pi$, and $3\pi/2$ are used.
We also set the initial phases of the vertical oscillations randomly. 
The initial epicycle amplitude of the stars is based on the Rayleigh distribution, for which $\langle \tilde{a}_{\rm{e, 0}}^2\rangle^{1/2}=\{0.125, 0.500\}$.
The initial vertical amplitude is fixed as $\tilde{a}_{\rm v,0}=0.125$.
We set the impact parameter uniformly between $0.5\leq\tilde{b}_{\rm 0}\leq6.5$.
Here, we assume that these parameters do not change before the initial condition: $\tilde{a}_{\rm e, ini}=\tilde{a}_{\rm e,0}$, $\tilde{a}_{\rm v, ini}=\tilde{a}_{\rm v,0}$, $\phi_{\rm e,ini}=\phi_{\rm e,0}, \phi_{\rm v,ini}=\phi_{\rm v,0}$, and $\tilde{b}_{\rm ini}=\tilde{b}_0$.
Based on the above, we set the initial position to be $32.0\leq-\alpha\tilde{b}_{\rm ini}\tilde{t}_{\rm ini}\leq160$.
To compare with the solution of the perturbation analysis (Appendix \ref{app:PA}), we prepare the same initial conditions for the perturbation analysis, but the vertical oscillation is neglected in this analysis.
The integrating time of the simulation is $\tilde{t}=140$.

The above initial conditions are summarized in Table \ref{tab:initial-condition}.

\begin{deluxetable}{ccccccc}
\tablecaption{The disk parameters and the initial condition of the orbital elements \label{tab:initial-condition}}
\tablehead{
	\colhead{model} &
	\colhead{$\Gamma$} &
	\colhead{$\tilde{b}_{0}$} & \colhead{$\tilde{a}_{\rm e, ini}$} & \colhead{$\tilde{a}_{\rm v,ini}$} & \colhead{$\phi_{\rm e,ini}$} & \colhead{$\phi_{\rm v,ini}$}
}
\startdata
			1 & 1.034 & 3.0 & 0.125, 0.500 & 0.125, 0.500 & 0,$\pi/32,\cdots,63\pi/32$ & 0 \\
			2 & 1.034 & 3.0 & 0.125 & 0.125 & 0,$\pi/8,\cdots,15\pi/8$ & $0,\pi/8,\cdots,15\pi/8$ \\
			3 & 1.034 & 2.5--6.0 & 0 & 0.125 & --- & 0 \\
			4 & 1.034 & 3.0, 4.0, 5.0, 6.0 & 0--2 & 0.125& 0,$\pi/32,\cdots,63\pi/32$ & 0 \\
			5 & 0.0625--1.5 & 3.0 & 0.125 & 0.125& 0,$\pi/32,\cdots,63\pi/32$ & 0 \\
			6 & 0.5, 1.0, 1.5 & 0.5--6.5 & Rayleigh (RMS = 0.125, 0.500) & 0.125 & random & random \\
			7 & 1.034 & 0.5--6.5 & Rayleigh (RMS = 0.125)& 0.125 & $0,\pi/2,\pi,3\pi/2$ & random \\
			8 & 1.034 & 0.5--6.5 & Rayleigh (RMS = 0.125) & --- & random & --- \\
			9 & 1.034 & 0.5--6.5 & Rayleigh (RMS = 0.125) & 0.125 & random & random 
\enddata
\end{deluxetable}

\section{Results}  \label{sec:result}

The change in the stellar orbits due to gravitational scattering by a perturber is investigated by numerical integration.
The orbital elements with subscripts ``fin" are the final values after gravitational scattering.
We vary the initial orbital elements $\tilde{b}_{0}, \tilde{a}_{\rm{e, ini}}, \tilde{a}_{\rm{v, ini}},\phi_{\rm e,ini},\phi_{\rm v,ini}$ and the disk parameter  $\Gamma$ to clarify their effects.

\subsection{Phase Synchronization of Epicyclic Motion} \label{sub:synchro}

\subsubsection{Phase Synchronization} \label{subsub:phase-synchro}

\begin{figure}
	\figurenum{1}
	\plotone{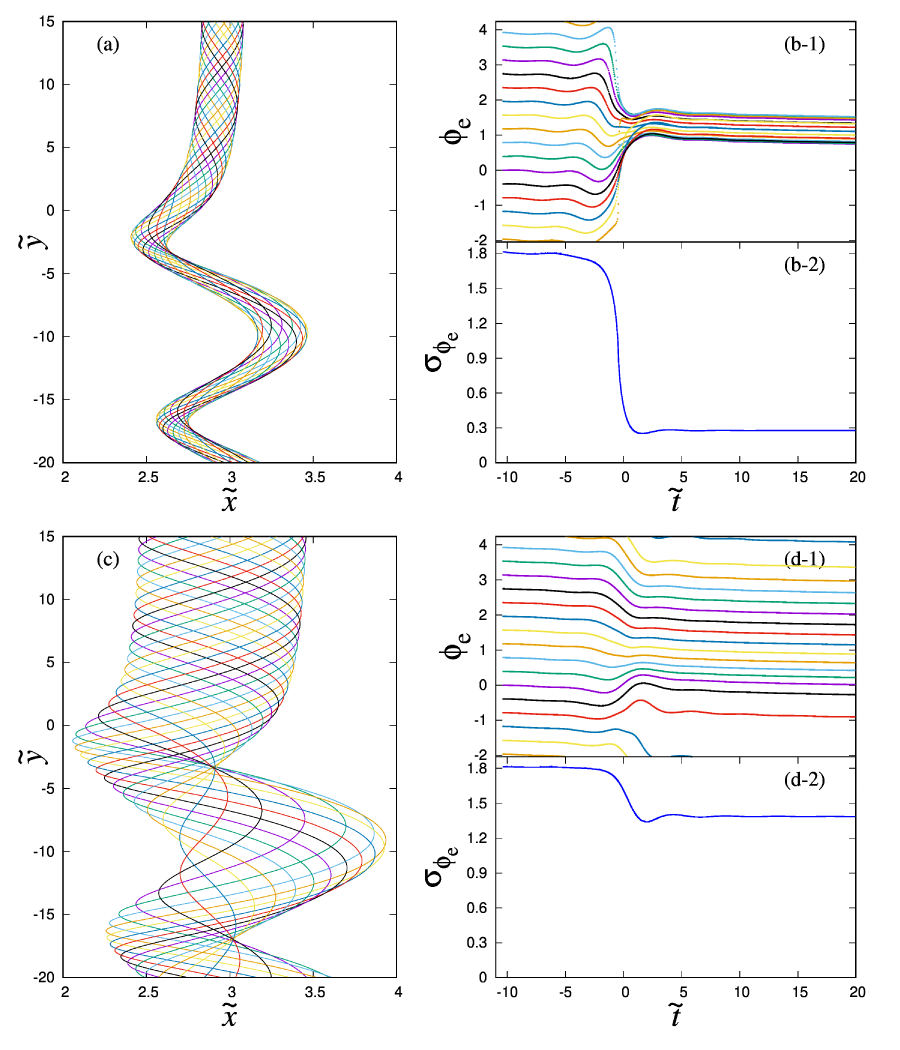}
	\caption{The right panels show the time evolution of the epicycle phases and the left panels show the trajectories of the stars.
	The different color lines correspond to the different initial epicycle phases.
	The top panels are the low velocity case and the bottom panels are the high velocity case. The perturber is placed at $(0,0)$. 
	When $\tilde{t}\simeq0$, the stars are at $\tilde{y}\simeq0$ and the gravity of the perturber is the strongest so that the scattering works well. 
	The synchronization of the epicyclic motion can be observed from the trajectories.\label{fig:time}}
\end{figure}

First, we demonstrate the dependence of the evolution of the epicycle phase on the epicycle amplitude.
The initial condition is model 1 and the initial vertical amplitude and phase are fixed as $\tilde{a}_{\rm v,ini}=0.125$ and $\phi_{\rm v,ini}=0$.
Figure \ref{fig:time} shows the stellar trajectories and the time evolution of  the epicycle phase and its deviation for $\tilde{a}_{\rm e, ini}=0.125$ and 0.500. 
For the small  $\tilde{a}_{\rm e, ini}$  case (top panels) , although the stars have different epicycle phases initially, their phases are synchronized after scattering (Figure \ref{fig:time}a, $\tilde{y}<0$ region).
The epicycle amplitude becomes $\tilde{a}_{\rm e,fin}=0.35\pm0.09$ and the impact parameter becomes $\tilde{b}_{\rm fin}=3.00\pm0.02$.
Generally the epicycle amplitude and the impact parameter increase by scattering for small $\tilde{a}_{\rm e, ini}$.
The epicycle phases converge to 1.1 and the standard deviation is around 0.28 (Figure \ref{fig:time}b). 
In other words, the gravity of the perturber causes the epicycle phases to be synchronized.

In contrast, the epicycle phases do not change significantly for large $\tilde{a}_{\rm e, ini}$  (bottom panels). 
The trajectories (Figure \ref{fig:time}c) show that the epicyclic motions after scattering are not the same.
The final epicycle amplitude and impact parameter are $\tilde{a}_{\rm e,fin}=0.55\pm0.23$ and $\tilde{b}_{\rm fin}=3.02\pm0.08$, respectively.
Generally the average of the epicycle amplitude and impact parameter increase less as $\tilde{a}_{\rm e,ini}$ increases.
In other words, the epicyclic motions barely change, so that phase synchronization does not occur (Figure \ref{fig:time}d). 

The above results clearly show that the degree of phase synchronization depends on the initial epicycle amplitude. 
We investigate this dependence in Section \ref{subsub:e_a}.

\subsubsection{Effect of Vertical Motion} \label{subsub:vertical}

\begin{figure}[ht!]
	\figurenum{2}
	\plotone{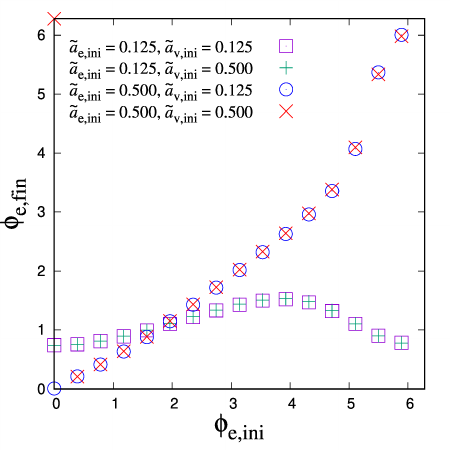}
	\caption{Dependence of the final epicycle phase on the initial epicycle phase for each initial condition.
	The open square and plus symbol data points correspond to the cases $\tilde{a}_{\rm v, ini}=0.125$, 0.500 and $\tilde{a}_{\rm e, ini}=0.125$ for each model.
	The open circle and cross symbol data points correspond to the cases $\tilde{a}_{\rm v, ini}=0.125$, 0.500 and $\tilde{a}_{\rm e, ini}=0.500$.
	The other initial parameters are $\tilde{b}_{0}=3.0$ and $\phi_{\rm v,ini}=0$, and the disk parameter is $\Gamma=1.034$.
	\label{fig:t-t}}
\end{figure}

We use model 1 to investigate the effect of the vertical motion on the phase synchronization.
Figure \ref{fig:t-t} shows the final epicycle phase against the initial epicycle phase for model 1.
The final phases for $\tilde{a}_{\rm e,ini}=0.125$ and $\tilde{a}_{\rm v, ini}=0.125$ and those for $\tilde{a}_{\rm e,ini}=0.125$ and $\tilde{a}_{\rm v, ini}=0.500$  overlap, and those for $\tilde{a}_{\rm e,ini}=0.500$ and $\tilde{a}_{\rm v, ini}=0.125$ and $\tilde{a}_{\rm e,ini}=0.500$ and $\tilde{a}_{\rm v, ini}=0.500$ overlap.
We find that the final epicycle phases are the same despite the different initial vertical amplitudes.

\begin{figure}[ht!]
	\figurenum{3}
	\plotone{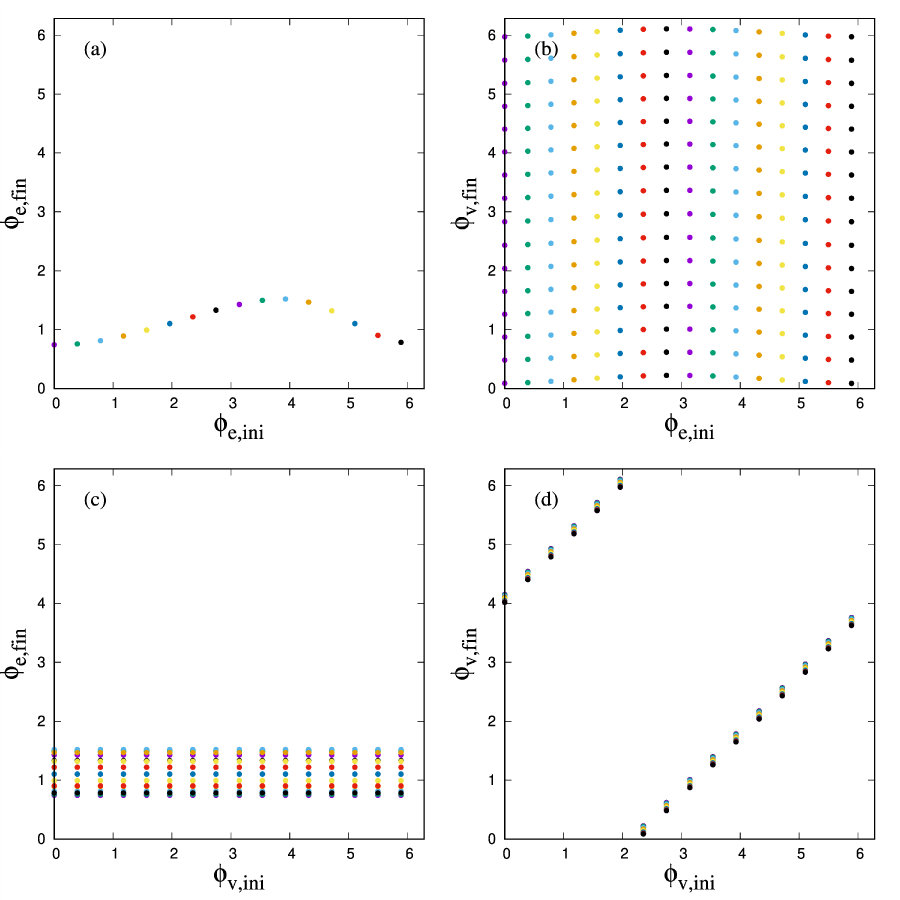}
	\caption{Final epicycle and vertical phases represented as a function of the initial epicycle and vertical phases.
	The different color points correspond to the different initial epicycle phases.
	The initial condition is model 2. Each panel shows 256 points. 
	(a) $\phi_{\rm e,ini}-\phi_{\rm e,fin}$: Each of the 16 data points overlap at each plotted point. 
	(b) $\phi_{\rm e,ini}-\phi_{\rm v,fin}$: Showing that the vertical phase is independent of the initial epicycle phase.	
	(c) $\phi_{\rm v,ini}-\phi_{\rm e,fin}$: A vertical difference occurs due to the initial epicycle phase. 
	(d) $\phi_{\rm v,ini}-\phi_{\rm v,fin}$: The vertical phases change uniformly.
	\label{fig:phase}}
\end{figure}

Next, we investigate the effect of the initial epicycle and vertical phases. 
Model 2 is used for the initial condition.
We use 16 initial phases for the epicyclic and vertical oscillations.
Figure \ref{fig:phase} shows the relation between the initial and final phases. As there are 16 patterns for each phase, each panel in the figure includes 256 data points.
The left panels show that the final epicycle phase is independent of the initial vertical phase. 
In panel (a), 16 data points overlap at each plotted point, which implies that the difference between the initial vertical phases does not affect the final epicycle phase. 
Panel (c) shows that the points are aligned vertically, which results from the difference in the initial epicycle phases. 
The final epicycle phase does not change even if the initial vertical phase varies. 
The right panels show the dependence of the final vertical phase on the initial phases. 
Panel (b) shows that the final vertical phase is almost independent of the initial epicycle phase.
On the other hand, in panel (d) the final vertical phase is strongly related to the initial vertical phase. 
However, the change in the vertical phases is the same for different initial vertical phases. 
The final vertical phases have no particular phases. 
These results confirm that the epicyclic motion is independent of the vertical motion. 
This behavior agrees with \cite{2014ApJ...787..174M}.
Thus, hereafter we use $\tilde{a}_{\rm v, ini}=0.125$ and $\phi_{\rm v,ini}=0$ in all models. 

\subsection{Dependence on Orbital Elements and a Disk Parameter} \label{sub:depend}

\subsubsection{Impact Parameter} \label{subsub:b}

Figure \ref{fig:time} shows that the gravity of the perturber is responsible for the synchronization of the epicycle phase.
We investigate the dependence of the final epicycle phase on the initial impact parameter that determines the strength of the gravity of the perturber. 
The impact parameter $\tilde{b}_{0}$  ranges form 2.5 to 6.0 (model 3).
For simplicity the initial epicycle amplitude is set to $\tilde{a}_{\rm e,ini}=0$. 

\begin{figure}[ht!]
	\figurenum{4}
	\plotone{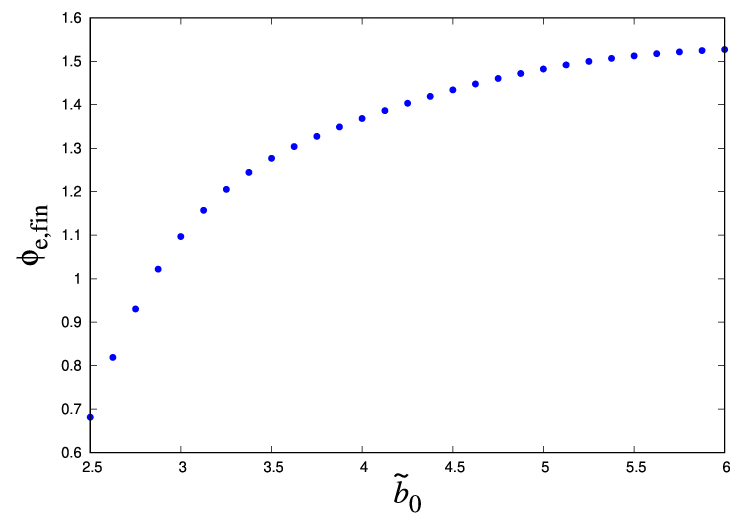}
	\caption{Dependence of the final epicycle phase on $\tilde{b}_{0}$. The initial condition is $\tilde{a}_{\rm e, ini}=0$, $\tilde{a}_{\rm v, ini}=0.125$ and $\phi_{\rm v,ini}=0$, and the disk parameter is $\Gamma=1.034$.
	\label{fig:b}}
\end{figure}

Figure \ref{fig:b} shows the final epicycle phase as a function of $\tilde{b}_{0}$. 
The final epicycle phase increases to around 1.5 with $\tilde{b}_{0}$.
For large $\tilde{b}_{0}$, the epicycle phase converges to $\pi/2$. 
The phase $\pi/2$ corresponds to the star moving towards the perturber at $\tilde{y}=0$. 
This value can be derived by perturbation analysis (Appendix \ref{app:PA}).

\subsubsection{Epicycle Amplitude} \label{subsub:e_a}

\begin{figure}[ht!]
	\figurenum{5}
	\plotone{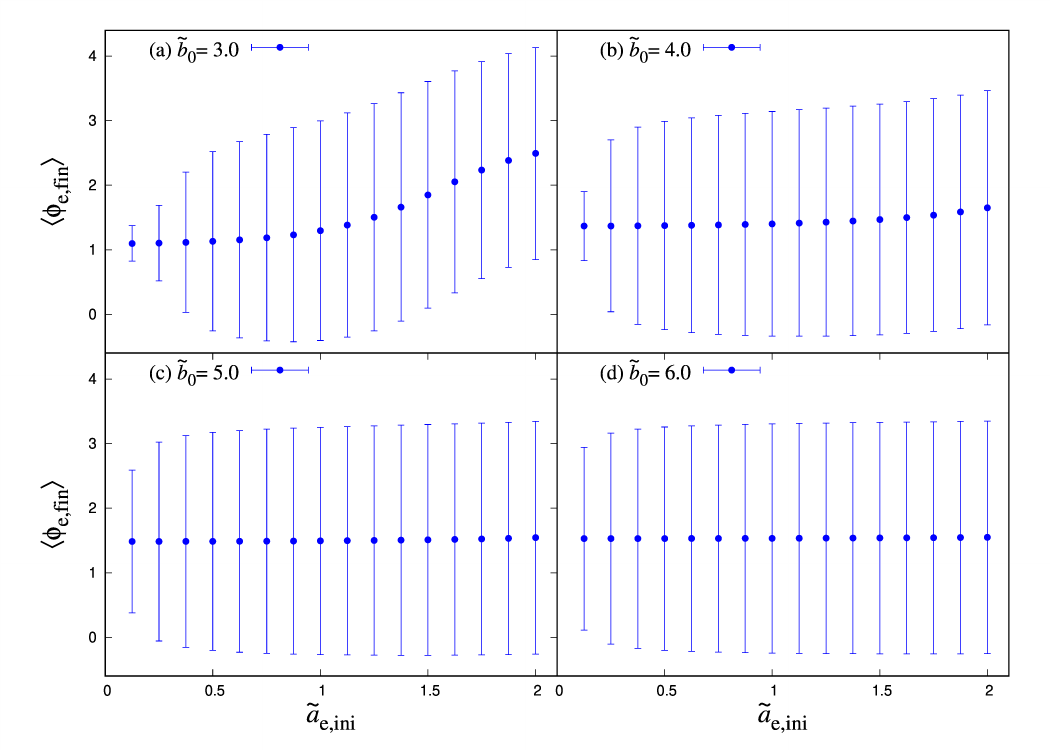}
	\caption{Dependence of the final epicycle phase on the initial epicycle amplitude for $\tilde{b}_{0}=3.0, 4.0, 5.0$, and $6.0$ in model 4. 
	The error bars represent the standard deviation.
	The deviation decreases as $\tilde{a}_{\rm e,ini}$ decreases.
	\label{fig:ea-vary}}
\end{figure}

We investigate the effect of the initial epicycle amplitude on the final epicycle phase.
The initial parameters are $\tilde{a}_{\rm e, ini}=(0,2.0]$ and  $\tilde{b}_{0}=\{3.0, 4.0, 5.0, 6.0\}$  (model 4). 
To calculate the standard deviation of the phases, an epicycle vector $\tilde{\bm{a}}_{\rm e}=(\tilde{a}_{\rm e}\cos\phi_{{\rm e}}, \tilde{a}_{\rm e}\sin\phi_{{\rm e}})$ is introduced. 
The average phase $\langle\phi_{\rm e}\rangle$ is defined as the argument of the average epicycle vectors, $\langle \tilde{\bm{a}}_{\rm e} \rangle$.
Figure \ref{fig:ea-vary} shows the average and standard deviation of the final epicycle phases. 
The large initial epicycle amplitude results in a large deviation, in other words, weak synchronization. 
For $\tilde{b}_{0}=3.0$, the average phase increases  with $\tilde{a}_{\rm e,ini}$, and for $\tilde{b}_{0}=4.0$, it increases slightly only for a large initial epicycle amplitude.  
On the other hand, the average phases for $\tilde{b}_{0}=5.0$ and 6.0 barely change over all $\tilde{a}_{\rm e, ini}$.
Figure \ref{fig:ea-vary} also shows that the average phase for $\tilde{a}_{\rm e,ini}\to0$ increases with $\tilde{b}_{0}$, and it agrees with the results of Figure \ref{fig:b}.

\begin{figure}[ht!]
	\figurenum{6}
	\plotone{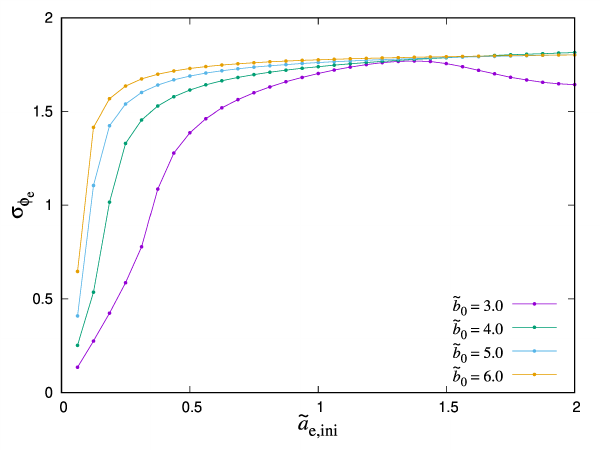}
	\caption{Standard deviation of the final epicycle phases as a function of the initial epicycle amplitude. 
	We use model 4 as the initial condition.
	The deviation decreases as $\tilde{b}_{0}$ and $\tilde{a}_{\rm e,ini}$ decrease.
	\label{fig:b-vary-devi}}
\end{figure}

Figure \ref{fig:b-vary-devi} shows the standard deviation as a function of $\tilde{a}_{\rm e,ini}$ for each $\tilde{b}_{0}$. 
For the same initial epicycle amplitude, the deviation increases with $\tilde{b}_{0}$.
This is because large $\tilde{b}_{0}$ leads to a large distance between the star and the perturber and thus their weak gravitational interaction.
For $\tilde{b}_{0}=4.0, 5.0$, and $6.0$, the deviation is saturated around $\sigma_{\phi_{\rm e}}\simeq1.8$ for $\tilde{a}_{\rm e,ini}>0.5$.
This corresponds to a random distribution of the final epicycle phases.
When the distribution of the phase is random, the deviation is equal to $\pi/\sqrt{3}\simeq1.81$.
Note that for a completely random distribution, the average vector $\tilde{\bm{a}}_{\rm e}$ is zero and thus the phase is not defined.
For $\tilde{b}_{0}=3.0$, the deviation of the epicycle phases is slightly smaller than that between $\tilde{a}_{\rm e,ini} = 1.5$ and 2.0.
This is because for large $\tilde{a}_{\rm e,ini}$ some stars are sufficiently close to the tidal radius of the perturber to be scattered strongly, and their epicycle phases are biased.
We find that phase synchronization needs a small $\tilde{b}_{0}$ and $\tilde{a}_{\rm e,ini}$. 

\subsubsection{Shear Rate} \label{subsub:gamma}

\begin{figure}[ht!]
	\figurenum{7}
	\plotone{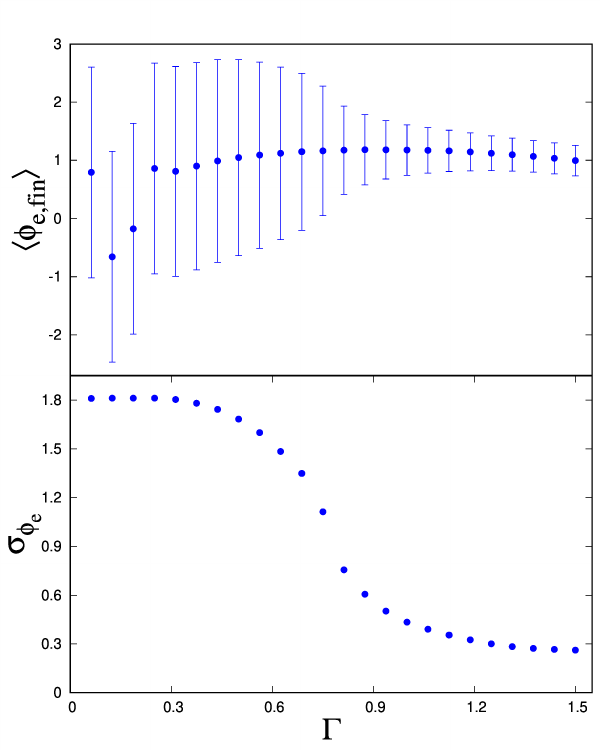}
	\caption{Final epicycle phases as a function of shear rate.
	The initial condition is model 5.
	The error bars represent the standard deviation.
	The deviation decreases rapidly between the shear rates of 0.5 and 1.0.
	\label{fig:shear-vary}}
\end{figure}

To investigate the dependence on the shear rate, we use model 5.
Figure \ref{fig:shear-vary} shows the dependence on $\Gamma$ of the averaged final epicycle phase and its standard deviation. 
The upper panel shows that the average phase is around 1.0 for $\Gamma\gtrsim0.4$, while it varies for small $\Gamma$.
This is because the average of the epicycle vectors is nearly a zero vector so that the average of the epicycle phases changes easily.
We find that a large shear rate results in a small deviation. 
Around $\Gamma\simeq0.7$, the deviation decreases rapidly with increasing $\Gamma$.
As the shear rate decreases, the standard deviation becomes large and saturates.
This is because a small shear rate leads to weak gravity in Equation (\ref{eq:normalized-motion}), and thus the epicycle phases are not synchronized. 
When the effect of gravity is large, the epicycle phases are well synchronized. 
However, a larger shear rate leads to a shorter timescale of the gravitational interaction.
We note that a large gravitational effect and a short timescale of the gravitational interaction have opposite roles for phase synchronization.
Figure \ref{fig:shear-vary} shows that a large shear rate is preferable for phase synchronization.

\subsection{Spatial Structure} \label{sub:density}

\subsubsection{Trailing Pattern}
\label{subsub:trailing-pattern}

We focus on Figure \ref{fig:map}b and investigate the structure around the perturber.
Figure \ref{fig:map}b shows how the surface density is amplified for the $\langle \tilde{a}^2_{\rm e,ini}\rangle^{1/2}=0.125$ and $\Gamma=1.0$ case, and we find trailing patterns around the perturber.
These trailing patterns are slightly curved
and the curves correspond to the orbits of the stars.
We find that the shear and epicyclic motion of the stars produce the trailing patterns in this shape.
Here, we focus on the first trailing pattern, although the amplification of the second trailing pattern is larger than that of the first, because the second trailing pattern is likely changed by self-gravity.
The surface density is most greatly amplified by 1.93 times at $(\tilde{x},\tilde{y})\simeq(3.2,-7)$ in first trailing pattern.
We find that gravitational scattering by the perturber creates the trailing patterns around the perturber.

\subsubsection{Dependence on Epicycle Parameters and Shear Rate} \label{subsub:random-map}

\begin{figure}
	\figurenum{8}
	\plotone{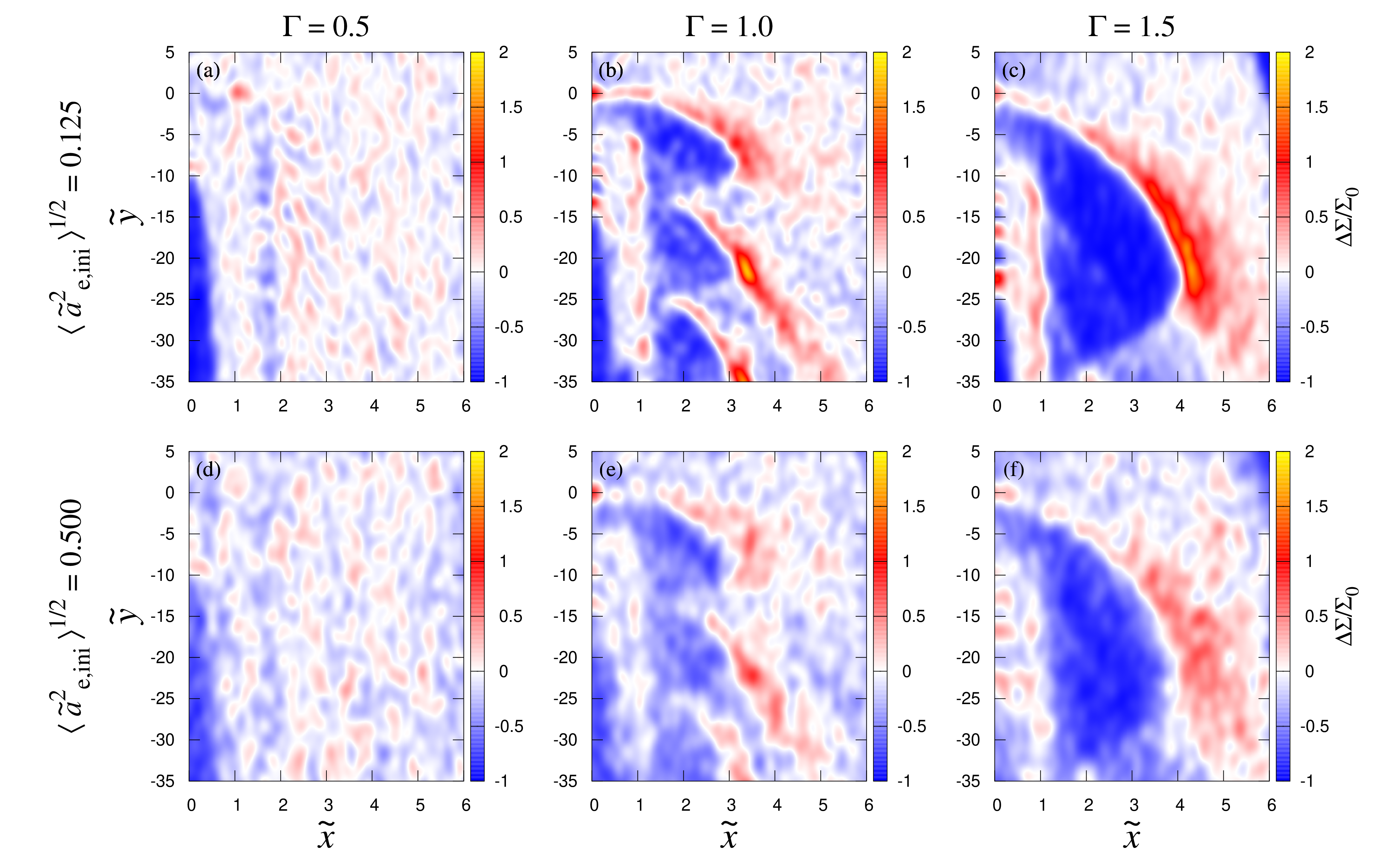}
	\caption{
	Density distribution of models of $\langle \tilde{a}_{\rm e,ini}^2\rangle^{1/2} =0.125$ (top panels) and $\langle \tilde{a}_{\rm e,ini}^2\rangle^{1/2} =0.500$ (bottom panels). 
	The disk parameter $\Gamma$ is 0.5, 1.0, and 1.5 from the left to right panels.
	\label{fig:map}}
\end{figure}

We investigate the dependence of the initial epicycle amplitude and the shear rate on the trailing patterns by using model 6.
Figure \ref{fig:map} shows the amplification distribution of the surface density around the perturber for each case.
For the $\Gamma=1.0$ and 1.5 cases, some trailing patterns are observed, while there are no substantial structures for the $\Gamma=0.5$ case.
The surface density is most greatly enhanced by 2.59 times for the $\langle \tilde{a}_{\rm e,ini}^2\rangle^{1/2}=0.125$ and $\Gamma=1.5$ case.
We find that the trailing pattern gets larger and its density enhancement increases with shear rate.
On the other hand, the initial epicycle amplitude does not affect the basic spatial pattern, while the density amplification weakens as it increases.

The location of the densest region depends on the shear rate. 
The trailing pattern becomes more parallel to $\tilde{y}$-axis for larger $\Gamma$.
The densest regions are located at $(\tilde{x},\tilde{y})\simeq(3.2, -7)$ for $\Gamma=1.0$ and $(\tilde{x},\tilde{y})\simeq(4.2, -20)$ for $\Gamma=1.5$.
For $\Gamma = 1.0$, the second and third trailing patterns are produced for $\tilde{y} > -35$. 
The second trailing pattern appears around $\tilde{y} \simeq -57$  for $\Gamma=1.5$. 

\added{Using the angle between the azimuthal direction and the line connecting the coordinate origin and the densest region as the proxy for the pitch angle of spiral arms $i$, we compare our results with the previous theoretical and observational studies.
We obtain $i \simeq 25^{\circ}$ for $\Gamma=1.0$ and $i \simeq 12^{\circ}$ for $\Gamma=1.5$.
These results are consistent with not only the theoretical model of spiral arms where $i = 22.0^{\circ}$ for $\Gamma=1.0$ and $i = 10.8^{\circ}$ for $\Gamma=1.5$ \citep{2014ApJ...787..174M} but also the observations where $i = 27.62\pm3.99^{\circ}$ for $\Gamma=1.0$ and $i = 9.32\pm5.05^{\circ}$ for $\Gamma=1.5$ \citep{2006ApJ...645.1012S,2014ApJ...795...90S}. 
It should be noted that the definition of the proxy pitch angle here is not rigorous since the trailing pattern is curved and the stellar self-gravity may affect the spatial structure as discussed in Section \ref{sub:self-gravity}.}

\begin{figure}
	\figurenum{9}
	\plotone{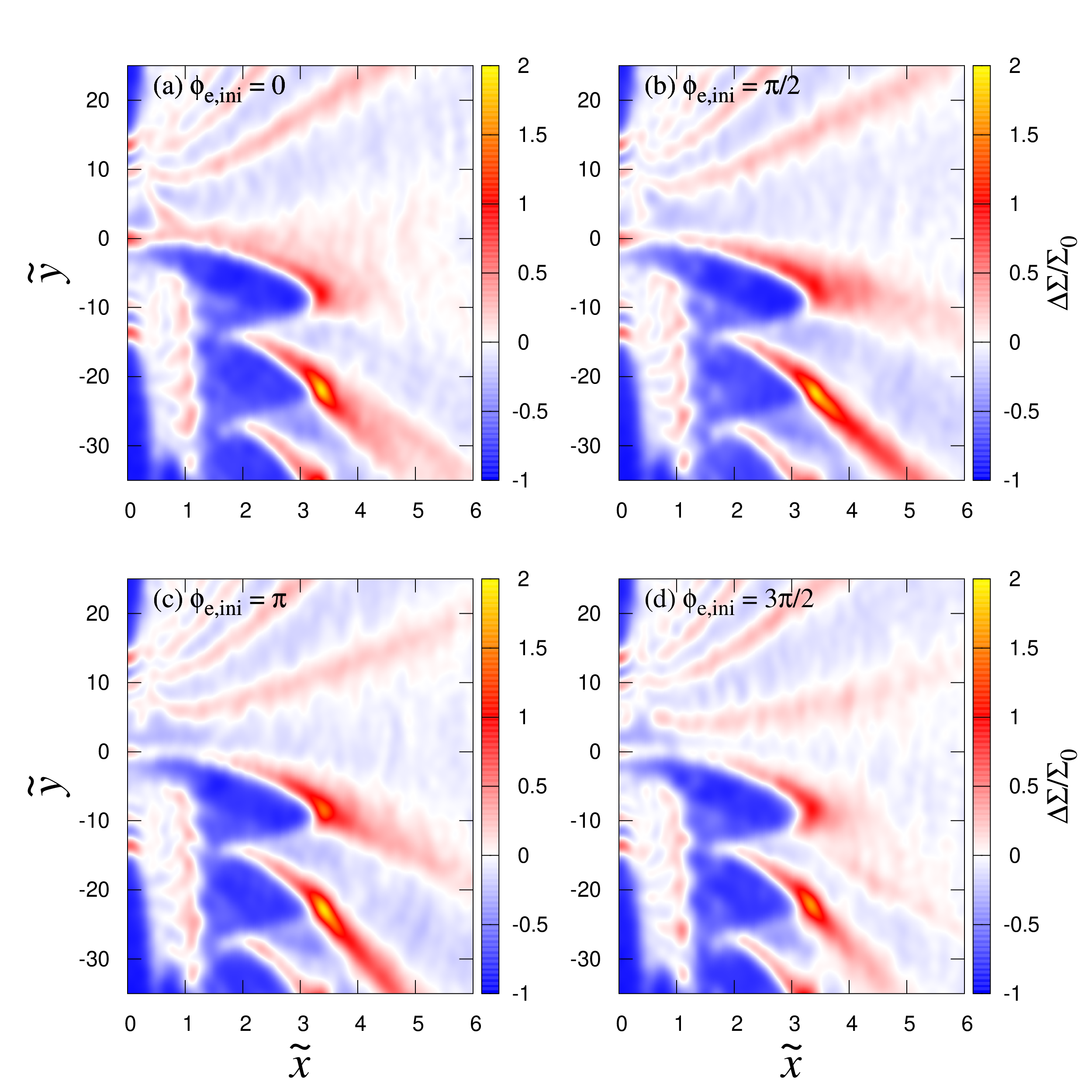}
	\caption{Snap shot of the density distribution. 
	The initial conditions are $\phi_{\rm e,ini}=0,\pi/2,\pi,3\pi/2$ (as indicated in each panel) and $\langle \tilde{a}^2_{\rm e,ini}\rangle^{1/2}=0.125$.
	\label{fig:leading}}
\end{figure}

Next, we take particular initial epicycle phases to investigate the effect on the spiral structure: $\phi_{\rm e,ini}=0,\pi/2,\pi,3\pi/2$ (model 7).
Figure \ref{fig:leading} shows the degree of amplification for each case.
The surface density is most amplified for the $\phi_{\rm e,ini}=\pi$ case (Figure \ref{fig:leading}c) and least amplified for the  $\phi_{\rm e,ini}=3\pi/2$ case (Figure \ref{fig:leading}d).
By comparing Figure \ref{fig:map}b, it is found that the surface density is more amplified for the $\phi_{\rm e,ini}=\pi$ case (Figure \ref{fig:leading}c) than for the result in Figure \ref{fig:map}b.
The condition of the epicycle phase $\phi_{\rm e}=\pi$ is that the gravitational force, the Coriolis force, and the tidal force are the same direction when the guiding center is at $\tilde{y}=0$, so that the gravitational force works efficiently. 
The structures except for the density are the same as in Figure \ref{fig:map}b. 

\subsubsection{Comparison with Perturbation Analysis}

\begin{figure}
	\figurenum{10}
	\plotone{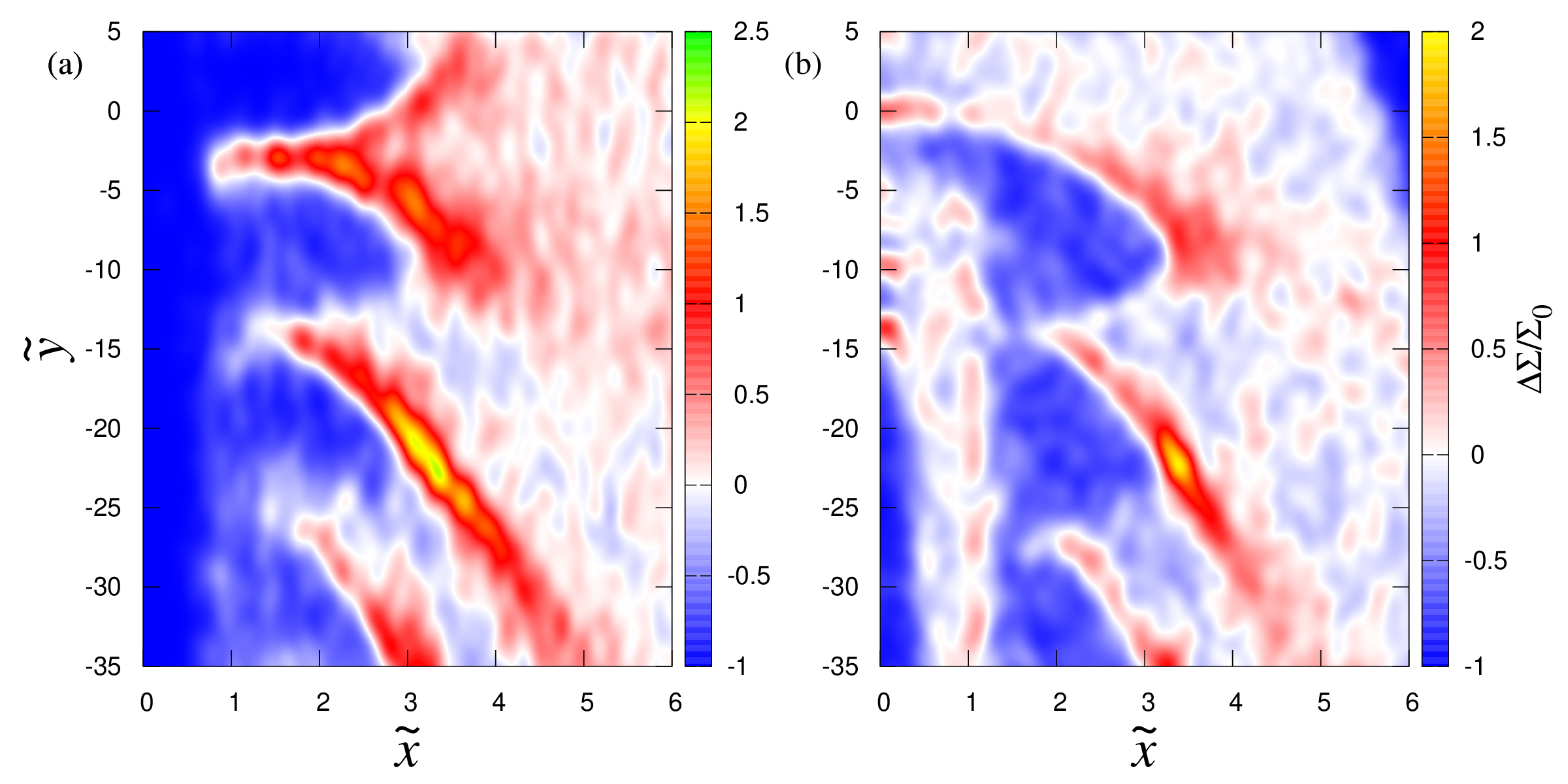}
	\caption{(a) Distribution of stars after scattering. 
	The positions of the stars are calculated by perturbation analysis; Appendix \ref{app:PA}. 
	(b) Snap shot of the distribution of stars. 
	The disk condition is $\gamma=1.39$ and $\beta=3.81$, and the initial condition is that the epicycle phases are random and $\langle \tilde{a}^2_{\rm e,ini}\rangle^{1/2}=0.125$ for each case.
	\label{fig:analy-simul}}
\end{figure}

We produce a distribution based on the perturbation analysis by using model 8.
Figure \ref{fig:analy-simul}a shows the distribution around the perturber.
In Figure \ref{fig:analy-simul}a, the surface density is the most amplified by a factor of 3.2.
To compare with the analysis, we produce a distribution in the same condition using model 9; Figure \ref{fig:analy-simul}b shows the result of the simulation. 
These two figures suggest that the structures are similar in these cases and that the perturbation analysis can reproduce the results of the simulation. 
However, the perturbation analysis does not include the change in the quadrant of the star's position, although some stars are observed to be strongly scattered and to move to the second quadrant in the simulation.
We find that the perturbation analysis is not rigorous, but can roughly evaluate the structures.

\section{Discussion} \label{sec:discussion}

\subsection{Toomre's Stability Criterion} \label{sub:q-value}

Toomre's stability criterion $Q$  is an indicator of the stability of differentially rotating stellar disks given by\citep{1964ApJ...139.1217T}
\begin{equation} \label{eq:q-value}
Q = \frac{\sigma_R \kappa}{3.36 G\Sigma_0},
\end{equation}
where $\Sigma_0$ is the mean surface density.
The critical wavelength of the gravitational instability is given by
\begin{equation} \label{eq:lambda-crit}
\lambda_{\rm{cr}}=\frac{4\pi^2G\Sigma_0}{\kappa^2}.
\end{equation}
The normalized wavelength can be calculated as
\begin{equation} \label{eq:normalized-lambda}
\tilde{\lambda}_{\rm{cr}} = \left(\frac{4\pi\alpha}{4-\alpha}\right)^{1/3},
\end{equation}
where the mass of the perturber is assumed to be $M=\pi\lambda_{\rm{cr}}^2\Sigma_0$. 
Toomre's $Q$  can be represented by normalized parameters as
\begin{equation} \label{eq:normalized-q}
Q = \frac{4\pi^2}{3.36}\frac{\tilde{\sigma}_R}{\gamma \tilde{\lambda}_{\rm cr}} \simeq \frac{5.1\tilde{\sigma}_R}{\alpha^{1/3}\gamma^{1/3}} \simeq 3.6\langle \tilde{a}_{\rm e}^2\rangle^{1/2} \frac{\gamma^{2/3}}{\alpha^{1/3}},
\end{equation}
where we use the relation between the velocity dispersion and the epicycle amplitude: $\sigma_{R}=\langle \dot{\tilde{x}}^2\rangle^{1/2} = \gamma \langle \tilde{a}_{\rm e}^2 \rangle^{1/2}/\sqrt{2}$.
For $Q=$ 1--2, spiral arms develop due to the self-gravity \citep{1964ApJ...140..646L,1981seng.proc..111T,1965MNRAS.130..125G,2016ApJ...821...35M}.
From Eq.~(\ref{eq:normalized-q}), we obtain $\langle \tilde{a}_{\rm e}^2 \rangle^{1/2}\simeq$ 0.3--0.6 for $Q=$ 1--2.
Figure \ref{fig:map} shows that a trailing dense region exists around the perturber.
If we consider the self-gravity, the amplitude of the region is likely to be more enhanced.
We propose that an elementary process of the swing amplification can be interpreted as gravitational scattering in differentially rotating disks.
The condition of the phase synchronization that  $\langle \tilde{a}_{\rm e}^2 \rangle^{1/2}$  should be small corresponds to the disk instability condition that $Q \lesssim 2$.

When the initial $Q$  is small, $Q$  increases rapidly to around 1.6 during swing amplification \citep{2014ApJ...787..174M,2018MNRAS.481..185M}.
This situation corresponds to an increase in the epicycle amplitude by scattering in our simulation.
This can be observed in Figure \ref{fig:time}a.
In contrast, \cite{2018MNRAS.481..185M} shows that $Q$ does not increase much when the initial $Q$ is large.
Figure \ref{fig:time}c corresponds to this situation.

\cite{2016ApJ...821...35M} showed that the parameters related to the spiral structure depend weakly on $Q$.
Figure \ref{fig:map} shows that the geometry of the structure is the same for different initial epicycle amplitude, which agrees with \cite{2016ApJ...821...35M}.

\subsection{Effect of Self-Gravity} \label{sub:self-gravity}

Our simulation does not include the self-gravity of stars. 
The self-gravity plays an important role in the swing amplification \citep{1981seng.proc..111T}.
It is considered that dense regions shrink and their density is more amplified, and that this trailing structure becomes firmer due to the self-gravity. 
Because the first region becomes denser and the trajectories of the stars are changed by the self-gravity, we cannot easily predict how the second region changes. 
An important feature of this paper is that the mechanism of the swing amplification is studied as the orbital evolution under the epicycle approximation.

\section{Summary} \label{sec:summary}

In order to understand the elementary process of swing amplification in spiral arm formation in disk galaxies we have investigated the phase synchronization of stellar orbits due to gravitational scattering by a perturber under the epicycle approximation. 
We confirmed phase synchronization of epicyclic motion by scattering and clarified its dependence on the orbital elements and the disk parameter, namely the impact parameter $\tilde{b}_{0}$, the epicycle amplitude $\tilde{a}_{\rm{e,ini}}$, the phase $\phi_{\rm e,ini}$, and the shear rate $\Gamma$. 
We calculated the standard deviation of the epicycle phases to evaluate the phase synchronization and also investigated the spatial structure formed by the phase synchronization.
The main findings in this paper can be summarized as follows:
\begin{itemize}
\item The smaller the initial epicycle amplitude, the smaller the impact parameter and the larger the shear rate, resulting in a smaller deviation of the epicycle phases. 
For $\tilde{a}_{\rm{e,ini}}\lesssim0.5$ and $\Gamma\gtrsim 0.8$, the epicycle phases are synchronized.
The vertical motion doesn't affect the phase synchronization.
\item In the surface density distribution of stars, the gravity of the perturber forms trailing dense regions by phase synchronization of the epicyclic motion.
For smaller $\langle \tilde{a}_{\rm e,ini}^2\rangle^{1/2}$ and larger $\Gamma$, the surface density is more amplified. 
For a larger shear rate, the structure is more parallel to the azimuthal direction, and its size is larger.
The group of stars with $\phi_{\rm e,ini}=\pi$  shows the densest trailing pattern.
\item Perturbation analysis reproduces well the spatial distribution obtained in the simulation.
\end{itemize}

In the present paper we demonstrated that gravitational scattering of stars by a softened point mass produces a trailing dense region, adding to the understanding of the basic physics.
However, in reality the perturber may not be a point mass but a finite-sized spiral arm.
In the next paper we will explore the effects of the perturber shape on the phase synchronization.

\acknowledgments

We wish to thank Shugo Michikoshi for advice regarding the relation between our results and that of {\em N}-body simulations.
E.K. is supported by JSPS KAKENHI Grant Number 18H05438.

\appendix

\section{Perturbation Analysis of Gravitational Scattering} \label{app:PA}

We introduce the new orbital elements $\tilde{a}_{{\rm e},j}(j=1,2)$ 
\begin{equation} \label{eq:intro-e-vector}
(\tilde{a}_{{\rm e},1}, \tilde{a}_{{\rm e},2}) = (\tilde{a}_{\rm e}\cos\phi_{\rm e}, \tilde{a}_{\rm e}\sin\phi_{\rm e}).
\end{equation}
From Equations (\ref{eq:solution}) and (\ref{eq:solution-velocity}), these elements satisfy
\begin{eqnarray} \label{eq:e1-and-e2}
\left\{
\begin{array}{l}
\displaystyle \tilde{a}_{{\rm e},1} = \frac{2\dot{\tilde{y}}+\alpha\tilde{x}}{\gamma^2}\cos\gamma \tilde{t} + \frac{\dot{\tilde{x}}}{\gamma}\sin\gamma \tilde{t},\\
\displaystyle \tilde{a}_{{\rm e},2} = \frac{2\dot{\tilde{y}}+\alpha\tilde{x}}{\gamma^2}\sin\gamma \tilde{t} - \frac{\dot{\tilde{x}}}{\gamma}\cos\gamma \tilde{t}.
\end{array}
\right.
\end{eqnarray}
From Equation (\ref{eq:normalized-motion}), the time derivative of Equation (\ref{eq:e1-and-e2}) can be given as
\begin{eqnarray} \label{eq:time-dif-e-vec}
\left\{
\begin{array}{l}
\displaystyle \dot{\tilde{a}}_{{\rm e},1} = \frac{2F_y}{\gamma^2} \cos \gamma \tilde{t} + \frac{F_x}{\gamma} \sin \gamma \tilde{t},\\
\displaystyle \dot{\tilde{a}}_{{\rm e},2} = \frac{2F_y}{\gamma^2} \sin \gamma \tilde{t} - \frac{F_x}{\gamma} \cos \gamma \tilde{t},
\end{array}
\right.
\end{eqnarray}
where $F_x$ and $F_y$ are the $x$- and $y$-components of the gravity from the perturber, given as ${\bm F}=-\alpha\tilde{r}^{-3}\bm{r}$. 
We adopt the impulse approximation where the stellar orbit is fixed and assume  $\varepsilon = \tilde{a}_{\rm e}/\tilde{b}\ll 1$.
Substituting  Eq.\ (\ref{eq:solution}) into $\bm r$ we expand $F_x$ and $F_y$  with $\varepsilon$ as
\begin{eqnarray} \label{eq:force}
\left\{
\begin{array}{l}
\displaystyle F_x = -\frac{\alpha}{\tilde{b}^2L^{3/2}} + \left[ \frac{\alpha}{\tilde{b}^2L^{3/2}} \cos(\gamma \tilde{t}-\phi_{\rm e}) - \frac{3\alpha}{\tilde{b}^2L^{5/2}}\left(\cos(\gamma \tilde{t}-\phi_{\rm e}) + \frac{\alpha}{\gamma}\sin(\gamma\tilde{t}-\phi_{\rm e})\right) \right] \varepsilon + \mathcal{O}(\varepsilon^2),\\
\displaystyle F_y = \frac{\alpha^2}{2}\frac{\tilde{t}}{\tilde{b}^2L^{3/2}} + \left[ \frac{3\alpha^2\tilde{t}}{2\tilde{b}^2 L^{5/2}} \left( \cos(\gamma\tilde{t}-\phi_{\rm e}) + \frac{\alpha}{\gamma} \sin(\gamma\tilde{t}-\phi_{\rm e}) \right) - \frac{2\alpha}{\gamma \tilde{b}^2 L^{3/2}} \sin(\gamma\tilde{t}-\phi_{\rm e})\right]\varepsilon + \mathcal{O}(\varepsilon^2),
\end{array}
\right.
\end{eqnarray}
where $L=1+\alpha^2\tilde{t}^2/4$.
The change in the new orbital elements by scattering is given by
\begin{eqnarray}
\left\{
\begin{array}{l}
\displaystyle \Delta \tilde{a}_{{\rm e},1} = \int_{-\infty}^{\infty} \dot{\tilde{a}}_{{\rm e},1}{\rm d}\tilde{t} = \frac{1}{\gamma \tilde{b}^2}[C_1 \sin \phi_{\rm e}\cdot \varepsilon + \mathcal{O}(\varepsilon^2)], \\
\displaystyle \Delta \tilde{a}_{{\rm e},2} = \int_{-\infty}^{\infty}\dot{\tilde{a}}_{\rm e,2}{\rm d}\tilde{t} = \frac{1}{\gamma \tilde{b}^2} \left[ C_2 + C_3 \cos\phi_{\rm e} \cdot \varepsilon + \mathcal{O}(\varepsilon^2) \right],
\end{array}\right.
\end{eqnarray}
where $C_1$, $C_2$, and $C_3$ are constants:
\begin{eqnarray} \label{eq:C}
\left\{
\begin{array}{l}
	\displaystyle C_1 = -2 + \frac{8}{\gamma^2}\left(\frac{4\gamma}{\alpha}\right)^2 K_0\left(\frac{4\gamma}{\alpha}\right) + \frac{8\gamma}{\alpha^2} (16-\alpha)K_1\left(\frac{4\gamma}{\alpha}\right) + 2\left(\frac{4\gamma}{\alpha}\right)^2 K_2\left(\frac{4\gamma}{\alpha}\right),\\
	\displaystyle C_2 = \frac{8\gamma}{\alpha}\left[\frac{2}{\gamma}K_0\left(\frac{2\gamma}{\alpha}\right) + K_1\left(\frac{2\gamma}{\alpha}\right)\right],\\
	\displaystyle C_3 = 2 + \frac{8}{\gamma^2}\left(\frac{4\gamma}{\alpha}\right)^2 K_0\left(\frac{4\gamma}{\alpha}\right) + \frac{8\gamma}{\alpha^2} (16-\alpha)K_1\left(\frac{4\gamma}{\alpha}\right) + 2\left(\frac{4\gamma}{\alpha}\right)^2 K_2\left(\frac{4\gamma}{\alpha}\right),
\end{array} \right.
\end{eqnarray}
where $K_j(x)$ are the modified Bessel functions.
Up to order $\mathcal{O}(\varepsilon)$ we obtain the final elements as
\begin{equation} \label{eq:fin-vec}
    \tilde{{\bm a}}_{\rm e,fin}= (\tilde{a}_{{\rm e},1} + \Delta \tilde{a}_{{\rm e},1}, \tilde{a}_{{\rm e},2} + \Delta \tilde{a}_{{\rm e},2}) \simeq (\Delta \tilde{a}_{{\rm e},1}, \Delta \tilde{a}_{{\rm e},2}) = \left( \mathcal{O}(\varepsilon),\  \frac{C_2}{\gamma \tilde{b}^2} + \mathcal{O}(\varepsilon) \right).
\end{equation}
Equation (\ref{eq:fin-vec}) shows that the final epicycle phase becomes around $\pi/2$ for small $\varepsilon$.

\bibliography{sample63}{}

\begin{thebibliography}{}
\expandafter\ifx\csname natexlab\endcsname\relax\def\natexlab#1{#1}\fi
\providecommand{\url}[1]{\href{#1}{#1}}
\providecommand{\dodoi}[1]{doi:~\href{http://doi.org/#1}{\nolinkurl{#1}}}
\providecommand{\doeprint}[1]{\href{http://ascl.net/#1}{\nolinkurl{http://ascl.net/#1}}}
\providecommand{\doarXiv}[1]{\href{https://arxiv.org/abs/#1}{\nolinkurl{https://arxiv.org/abs/#1}}}

\bibitem[{{Baba}(2015)}]{2015MNRAS.454.2954B}
{Baba}, J. 2015, \mnras, 454, 2954, \dodoi{10.1093/mnras/stv2220}

\bibitem[{{Baba} {et~al.}(2009){Baba}, {Asaki}, {Makino}, {Miyoshi}, {Saitoh},
  \& {Wada}}]{2009ApJ...706..471B}
{Baba}, J., {Asaki}, Y., {Makino}, J., {et~al.} 2009, \apj, 706, 471,
  \dodoi{10.1088/0004-637X/706/1/471}

\bibitem[{{Binney} \& {Tremaine}(1987)}]{1987gady.book.....B}
{Binney}, J., \& {Tremaine}, S. 1987, {Galactic dynamics} (Princeton University
  Press)

\bibitem[{{Fujii} {et~al.}(2011){Fujii}, {Baba}, {Saitoh}, {Makino}, {Kokubo},
  \& {Wada}}]{2011ApJ...730..109F}
{Fujii}, M.~S., {Baba}, J., {Saitoh}, T.~R., {et~al.} 2011, \apj, 730, 109,
  \dodoi{10.1088/0004-637X/730/2/109}

\bibitem[{{Goldreich} \& {Lynden-Bell}(1965)}]{1965MNRAS.130..125G}
{Goldreich}, P., \& {Lynden-Bell}, D. 1965, \mnras, 130, 125,
  \dodoi{10.1093/mnras/130.2.125}

\bibitem[{{Julian} \& {Toomre}(1966)}]{1966ApJ...146..810J}
{Julian}, W.~H., \& {Toomre}, A. 1966, \apj, 146, 810, \dodoi{10.1086/148957}

\bibitem[{{Kokubo} \& {Ida}(1992)}]{1992PASJ...44..601K}
{Kokubo}, E., \& {Ida}, S. 1992, \pasj, 44, 601

\bibitem[{{Lin} \& {Shu}(1964)}]{1964ApJ...140..646L}
{Lin}, C.~C., \& {Shu}, F.~H. 1964, \apj, 140, 646, \dodoi{10.1086/147955}

\bibitem[{{Michikoshi} \& {Kokubo}(2014)}]{2014ApJ...787..174M}
{Michikoshi}, S., \& {Kokubo}, E. 2014, \apj, 787, 174,
  \dodoi{10.1088/0004-637X/787/2/174}

\bibitem[{{Michikoshi} \& {Kokubo}(2016{\natexlab{a}})}]{2016ApJ...821...35M}
---. 2016{\natexlab{a}}, \apj, 821, 35, \dodoi{10.3847/0004-637X/821/1/35}

\bibitem[{{Michikoshi} \& {Kokubo}(2016{\natexlab{b}})}]{2016ApJ...823..121M}
---. 2016{\natexlab{b}}, \apj, 823, 121, \dodoi{10.3847/0004-637X/823/2/121}

\bibitem[{{Michikoshi} \& {Kokubo}(2018)}]{2018MNRAS.481..185M}
---. 2018, \mnras, 481, 185, \dodoi{10.1093/mnras/sty2274}

\bibitem[{{P{\'e}rez} {et~al.}(2016){P{\'e}rez}, {Carpenter}, {Andrews},
  {Ricci}, {Isella}, {Linz}, {Sargent}, {Wilner}, {Henning}, {Deller},
  {Chandler}, {Dullemond}, {Lazio}, {Menten}, {Corder}, {Storm}, {Testi},
  {Tazzari}, {Kwon}, {Calvet}, {Greaves}, {Harris}, \&
  {Mundy}}]{2016Sci...353.1519P}
{P{\'e}rez}, L.~M., {Carpenter}, J.~M., {Andrews}, S.~M., {et~al.} 2016,
  Science, 353, 1519, \dodoi{10.1126/science.aaf8296}

\bibitem[{{Pettitt} {et~al.}(2016){Pettitt}, {Tasker}, \&
  {Wadsley}}]{2016MNRAS.458.3990P}
{Pettitt}, A.~R., {Tasker}, E.~J., \& {Wadsley}, J.~W. 2016, \mnras, 458, 3990,
  \dodoi{10.1093/mnras/stw588}

\bibitem[{{Seigar} {et~al.}(2006){Seigar}, {Bullock}, {Barth}, \&
  {Ho}}]{2006ApJ...645.1012S}
{Seigar}, M.~S., {Bullock}, J.~S., {Barth}, A.~J., \& {Ho}, L.~C. 2006, \apj,
  645, 1012, \dodoi{10.1086/504463}

\bibitem[{{Seigar} {et~al.}(2014){Seigar}, {Davis}, {Berrier}, \&
  {Kennefick}}]{2014ApJ...795...90S}
{Seigar}, M.~S., {Davis}, B.~L., {Berrier}, J., \& {Kennefick}, D. 2014, \apj,
  795, 90, \dodoi{10.1088/0004-637X/795/1/90}

\bibitem[{{Sellwood}(2000)}]{2000Ap&SS.272...31S}
{Sellwood}, J.~A. 2000, \apss, 272, 31, \dodoi{10.1023/A:1002668818252}

\bibitem[{{Sellwood} \& {Carlberg}(1984)}]{1984ApJ...282...61S}
{Sellwood}, J.~A., \& {Carlberg}, R.~G. 1984, \apj, 282, 61,
  \dodoi{10.1086/162176}

\bibitem[{{Sellwood} \& {Lin}(1989)}]{1989MNRAS.240..991S}
{Sellwood}, J.~A., \& {Lin}, D.~N.~C. 1989, \mnras, 240, 991,
  \dodoi{10.1093/mnras/240.4.991}

\bibitem[{{Toomre}(1964)}]{1964ApJ...139.1217T}
{Toomre}, A. 1964, \apj, 139, 1217, \dodoi{10.1086/147861}

\bibitem[{{Toomre}(1981)}]{1981seng.proc..111T}
{Toomre}, A. 1981, in Structure and Evolution of Normal Galaxies, ed. S.~M.
  {Fall} \& D.~{Lynden-Bell}, 111--136

\bibitem[{{Yu} \& {Ho}(2019)}]{2019ApJ...871..194Y}
{Yu}, S.-Y., \& {Ho}, L.~C. 2019, \apj, 871, 194,
  \dodoi{10.3847/1538-4357/aaf895}

\end{thebibliography}
\bibliographystyle{aasjournal}

\end{document}